\documentclass[
  pra,
  twocolumn,
  superscriptaddress,
  amsmath,
  amssymb,
  aps,
  floatfix
]{revtex4-2}

\usepackage{graphicx}
\usepackage{subcaption}
\usepackage{bm}
\usepackage{braket}
\usepackage{booktabs}
\usepackage{xcolor}
\usepackage{ragged2e}
\usepackage[colorlinks=true,allcolors=blue]{hyperref}

\begin{document}

\title{Fisher-Information Recovery in Superconducting-Qubit Magnetometry with Squeezed-Microwave Readout}

\author{M.-R. Yun}
\affiliation{Institute of Quantum Materials and Physics, Henan Academy of Sciences, Zhengzhou 450046, China}
\author{Y.-J. Qu}
\affiliation{Department of Physics, Central China Normal University, Wuhan 430072, China}
\author{Zheng Shan}
\affiliation{State Key Laboratory of Mathematical Engineering and Advanced Computing, Zhengzhou, China}
\author{L.-L. Yan}
%\email{llyan@zzu.edu.cn}
\affiliation{Quantum Information Institute, School of Physics and Laboratory of Zhongyuan Light,
Zhengzhou University, Zhengzhou 450001, China}
 \affiliation{Institute of Quantum Materials and Physics, Henan Academy of Sciences, Zhengzhou 450046, China}
\author{Yu Jia}
\email{jiayu@zzu.edu.cn}
\affiliation{Institute of Quantum Materials and Physics, Henan Academy of Sciences, Zhengzhou 450046, China}
\author{S.-L. Su}
\email{slsu@zzu.edu.cn}
\affiliation{Quantum Information Institute, School of Physics and Laboratory of Zhongyuan Light,
Zhengzhou University, Zhengzhou 450001, China}
 \affiliation{Institute of Quantum Materials and Physics, Henan Academy of Sciences, Zhengzhou 450046, China}

\begin{abstract}
The performance of superconducting-qubit magnetometers depends not only on magnetic-field encoding during Ramsey interrogation, but also on how efficiently the encoded information is recovered during readout. Squeezed microwaves provide a promising route to improve information extraction by suppressing readout noise, yet how such readout enhancement translates into accessible information for magnetic-field estimation remains unclear. Here we develop an effective detected-mode framework linking projected quadrature noise, state-assignment error, and the classical Fisher information accessible from binary readout outcomes. A finite mismatch between the squeezed quadrature and the discrimination axis produces an optimal squeezing strength through the competition between squeezed and anti-squeezed fluctuations. We further show that squeezed readout improves the readout-limited magnetic-field sensitivity by recovering information otherwise lost during state assignment, with the enhancement remaining robust against quadrature mismatch, transmission loss, and added noise. These results establish a quantitative connection between squeezed-microwave readout and measurement-stage information recovery in superconducting quantum sensing.
\end{abstract}

\maketitle

\section{Introduction}
Quantum sensing exploits quantum systems to detect and estimate weak physical signals with high precision~\cite{RevModPhys.89.035002,Giovannetti2011,doi:10.1142/S0219749909004839,RevModPhys.90.035005,Hecht2025,Kristen2020,PhysRevLett.124.020501,PhysRevLett.126.010502}. Flux-sensitive superconducting circuits provide a natural platform for magnetometry and have been widely explored for quantum sensing and metrological applications~\cite{PhysRevA.76.042319,PhysRevB.77.180502,PhysRevApplied.8.044003,PhysRevB.105.024507,PhysRevB.110.214423,Kolosvetov2026,PhysRevA.107.052609,vq6z-3ms7,PhysRevLett.128.150501,Li2026,wb5z-y5y5}. In Ramsey magnetometry, magnetic flux shifts the qubit transition frequency, which is converted into measurable phase and population responses~\cite{ramsey1950molecular,Bal2012,Danilin2018}. The attainable precision therefore depends not only on magnetic-field encoding during Ramsey interrogation, but also on how efficiently the encoded information is recovered through the final measurement~\cite{PhysRevLett.72.3439,PhysRevApplied.7.054020}.

In circuit quantum electrodynamics, the qubit state is commonly inferred from the state-dependent dispersive response of a microwave resonator~\cite{RevModPhys.93.025005,PhysRevA.79.013819,PhysRevLett.105.100504,PhysRevLett.106.110502,PhysRevApplied.10.034040,PhysRevApplied.21.024008,PhysRevApplied.23.054057,1dg9-b4vw}. The two qubit states produce distinct resonator pointer states that are discriminated along a calibrated measurement quadrature~\cite{PhysRevA.77.012112,PhysRevLett.112.190504,Hacohen-Gourgy2016}. Vacuum fluctuations, amplifier noise, and other readout imperfections broaden the corresponding state-conditioned distributions, increasing their overlap and thereby producing state-assignment errors. From an estimation-theory perspective, the quantum Fisher information (QFI) sets the maximum information about the parameter available from the sensing state, whereas the classical Fisher information (CFI) quantifies the information contained in the outcomes of a specific measurement~\cite{PhysRevLett.72.3439,PhysRevA.90.014101}. Readout imperfections can therefore degrade the accessible CFI relative to ideal measurement, even after the magnetic-field dependence has been established during Ramsey interrogation~\cite{PhysRevA.95.012117}.

Microwave squeezing provides a direct route to reducing the quadrature noise that limits dispersive qubit readout. Squeezed microwave fields and near-quantum-limited parametric amplification have been realized in superconducting circuits~\cite{PhysRevLett.106.220502,PhysRevLett.109.250502,Bergeal2010,doi:10.1126/science.aaa8525,PhysRevApplied.9.044023}, enabling squeezing-assisted readout with improved signal-to-noise ratio, measurement rate, and state-assignment fidelity~\cite{PhysRevLett.115.093604,PhysRevLett.120.040505,PhysRevApplied.18.064092,PhysRevLett.129.123602,PhysRevLett.133.233605,Kam2024,xv37-xwmk,PhysRevX.7.041011,2s1m-y9bd}. For magnetometry, however, improvements in these readout metrics do not directly determine how efficiently the magnetic-field information encoded during Ramsey interrogation is recovered in the measurement outcomes. Establishing this connection requires relating the detected quadrature noise and resulting assignment error to the accessible Fisher information and, ultimately, to the magnetic-field sensitivity.

Here we develop an effective detected-mode framework for superconducting-qubit magnetometry with squeezed dispersive readout. The framework connects the detected quadrature noise to state-assignment error, the classical Fisher information accessible from binary measurement outcomes, and the resulting readout-limited magnetic-field sensitivity, while incorporating quadrature mismatch, transmission loss, and added measurement noise. We show that a finite quadrature mismatch produces an optimal squeezing strength through the competition between squeezed and anti-squeezed fluctuations, while the resulting information-recovery advantage remains robust against realistic readout imperfections and can approach the upper bound set by the vacuum-reference assignment error.

\section{MAGNETOMETER MODEL AND SENSING PROTOCOL}
\label{sec:magnetometer_model}
\begin{figure*}[t]
    \centering

    % (a) 上面的大图
    \includegraphics[width=\textwidth]{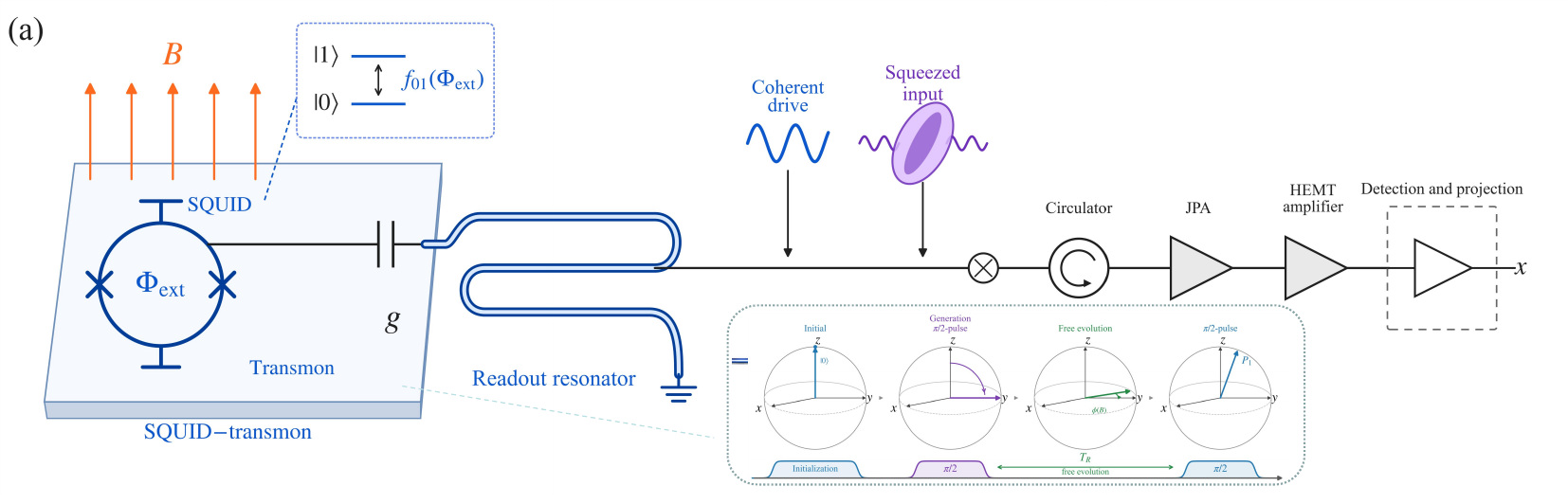}

    \vspace{2mm}

    % (b) 下面的大图
    \includegraphics[width=\textwidth]{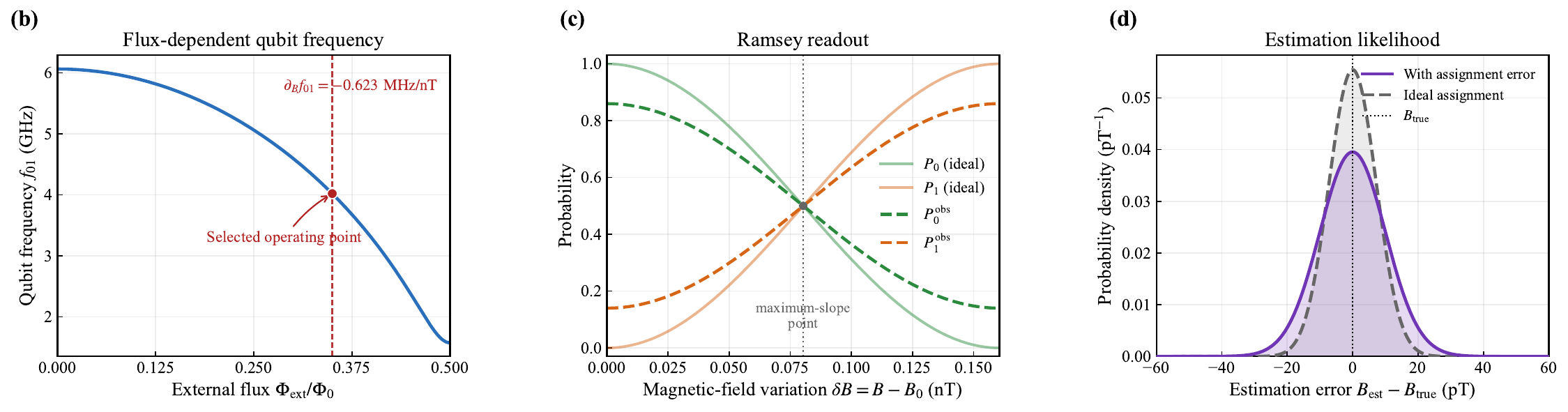}

 \caption{\justifying
Schematic and operating principle of superconducting-qubit
magnetometry with squeezed-microwave-assisted dispersive readout.
(a) Measurement architecture and Ramsey phase encoding.
(b) Flux-dependent qubit transition frequency with the selected operating point.
(c) Ideal and observed Ramsey populations with representative symmetric
state-assignment error.
(d) Corresponding finite-sample magnetic-field likelihoods, showing the
broadening induced by imperfect readout.
}
    \label{fig:scheme}
\end{figure*}
\subsection{Flux-sensitive transmon sensor}

We consider a flux-tunable transmon operated as a magnetic-field
sensor, as illustrated in Fig.~\ref{fig:scheme}(a). The two Josephson junctions form
a superconducting quantum interference device (SQUID), making the
qubit transition frequency sensitive to the magnetic flux threading
the loop. A magnetic field applied normal to the SQUID changes this flux through the effective loop area and thereby shifts the qubit transition frequency.

The device is operated around a static magnetic-field operating point $B_0$. Denoting the qubit transition frequency by \(f_{01}=\omega_q/2\pi\), for a sufficiently small field variation $\delta B=B-B_0$, the
resulting transition-frequency shift can be written in the linear
form
\begin{equation}
    \delta\omega_q \simeq G_B\,\delta B,
    \qquad
    G_B=
    \left.
    \frac{\partial\omega_q}{\partial B}
    \right|_{B_0},
    \label{eq:magnetic_transduction}
\end{equation}
where $G_B$ characterizes the local magnetic-field-to-frequency
transduction of the sensor. The corresponding flux-dependent transition
frequency is shown in
Fig.~\ref{fig:scheme}(b). We choose an operating point in a large-slope region to obtain a strong magnetic response while retaining a suitable qubit transition frequency.

This field-dependent frequency shift is converted into a relative qubit phase during the Ramsey interrogation described below. The SQUID-transmon Hamiltonian and the derivation of the flux-dependent transition frequency are given in Appendix~\ref{app:transmon}.

\subsection{Ramsey phase encoding}

The magnetic-field-induced frequency shift is converted into a relative qubit phase through a Ramsey sequence, whose state evolution on the Bloch sphere is illustrated in the inset of Fig.~\ref{fig:scheme}(a).
Starting from the ground state, a first $\pi/2$ pulse prepares an equal
superposition of the two qubit states. During a free-evolution
interval $T$, the field-dependent frequency shift produces the
relative phase
\begin{equation}
    \phi_B = G_B\,\delta B\,T .
    \label{eq:ramsey_phase}
\end{equation}
At this stage, the magnetic-field information is encoded in the
relative qubit phase.

A second $\pi/2$ pulse with analysis phase $\phi_a$ maps this phase
onto the qubit population. The excited-state probability after the
Ramsey sequence is
\begin{equation}
    p_1(B)
    =
    \frac{1-\cos\!\left(\phi_B+\phi_a\right)}{2}.
    \label{eq:ramsey_probability}
\end{equation}
The resulting Ramsey population response is shown in Fig.~\ref{fig:scheme}(c).
The analysis phase $\phi_a$ sets the operating point of the Ramsey
fringe.
The analysis phase is chosen such that the sensor operates near the
midfringe of the Ramsey oscillation, where the population response
to a small magnetic-field variation is maximal.

After the final analysis pulse, the field-dependent phase is encoded
in the qubit populations. In a circuit-QED implementation, these
populations are experimentally accessed through dispersive microwave
readout, as illustrated in Fig.~\ref{fig:scheme}. At this stage, we first treat the
final qubit measurement as an ideal binary measurement in order to
establish the magnetic-field estimation protocol. The physical
dispersive-readout process and its associated measurement noise are
introduced in Sec.~\ref{sec3}.

\subsection{Magnetic-field estimation from Ramsey populations}

The magnetic field is inferred from repeated measurements of the
final qubit state. Assuming ideal state discrimination, we consider
$N$ independent repetitions of the Ramsey sequence. If the excited
state is detected $n_1$ times, the corresponding binomial likelihood
for a trial field $B$ is
\begin{equation}
    \mathcal{L}(B|n_1,N)
    \propto
    [p_1(B)]^{n_1}
    [1-p_1(B)]^{N-n_1}.
    \label{eq:ideal_likelihood}
\end{equation}
The magnetic-field estimate $\hat B$ is obtained from the maximum of
this likelihood within the local sensing interval.

The width of the likelihood characterizes the finite-sample
uncertainty of the field estimate. Figure~\ref{fig:scheme}(d) illustrates the
corresponding likelihood distribution for ideal state assignment and
for a finite assignment error. When assignment errors are present, overlap between the state-conditioned readout distributions broadens the inferred-field likelihood. We next introduce the dispersive-readout model and quantify
how squeezed microwaves reduce this readout-induced information loss.

\section{Squeezed-microwave dispersive readout}\label{sec3}
We now describe how the Ramsey population is converted into binary
measurement outcomes and how squeezed microwaves modify the
readout noise. The analysis is formulated directly in terms of the
temporally integrated discrimination quadrature measured after the
state-dependent resonator response.

\subsection{State-dependent resonator response}
The qubit is dispersively coupled to a microwave resonator whose
response depends on the qubit state. A coherent readout pulse therefore
produces two distinct state-conditioned resonator responses. For a
readout pulse with envelope $\epsilon_d(t)$, the corresponding
resonator amplitudes satisfy
\begin{equation}
    \dot{\alpha}_j(t)
    =
    -\left[
    \frac{\kappa}{2}
    +i(\Delta+\chi_j)
    \right]\alpha_j(t)
    -i\epsilon_d(t),
    \qquad j=0,1 ,
    \label{eq:cavity_response}
\end{equation}
where $\kappa$ is the resonator linewidth, $\Delta$ is the readout
detuning, and $\chi_j$ is the state-dependent dispersive shift.

After temporal-mode integration and projection onto the calibrated
discrimination quadrature $x$, the two qubit states define projected
pointer-state means $m_0$ and $m_1$. Their separation,
\begin{equation}
    \Delta m=|m_1-m_0|,
    \label{eq:pointer_separation}
\end{equation}
sets the coherent signal available for state discrimination. The
input--output relation and the temporal-mode integration are given in
Appendix~\ref{app:readout}.

The incident fluctuations experience the same state-dependent
resonator response. For a narrowband single-port resonator, the
fluctuations near the readout frequency acquire the reflection
coefficient
\begin{equation}
    S_j
    =
    1-
    \frac{\kappa}
    {\kappa/2+i(\Delta+\chi_j)}
    \equiv e^{i\varphi_j},
    \label{eq:state_scattering}
\end{equation}
where $|S_j|=1$ in the lossless single-port limit. The resonator
therefore rotates the incident squeezed fluctuations by a
qubit-state-dependent phase $\varphi_j$. This phase rotation determines
the effective alignment between the squeezed quadrature and the
measurement quadrature considered below.

\begin{figure*}[t]
  \centering
  \includegraphics[width=\textwidth]{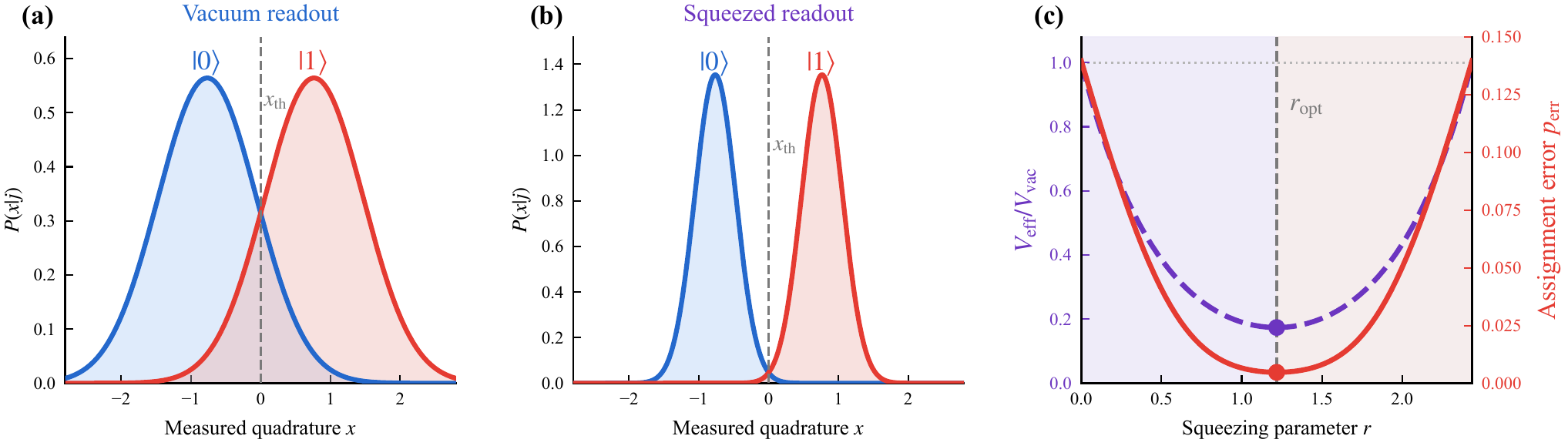}
  \caption{\justifying
Squeezing-enhanced qubit readout.
(a) State-conditioned distributions along the discrimination quadrature
$x$ for vacuum readout.
(b) Distributions at the optimal squeezing strength, showing reduced
overlap at fixed pointer-state separation.
(c) Normalized projected variance $V_{\rm eff}/V_{\rm vac}$ (purple
dashed) and assignment error $p_{\rm err}$ (red solid) versus squeezing
parameter $r$ for $\theta_{\rm mis}=5^\circ$. The vertical dashed line
indicates $r_{\rm opt}$.
}
  \label{fig:magnetometer}
\end{figure*}

\subsection{State-dependent squeezed-noise propagation}

A squeezed microwave field is applied during dispersive readout.
Squeezing redistributes fluctuations between orthogonal field
quadratures, suppressing noise along the squeezed quadrature while
amplifying fluctuations along its conjugate
quadrature
~\cite{PhysRevLett.60.764,PhysRevLett.117.020502,
PhysRevLett.119.023602}. 

Let $\phi_s$ denote the input squeezing angle and $\phi_m$ the
calibrated discrimination-quadrature angle. After the
state-dependent resonator rotation $\varphi_j$, the effective mismatch
between the squeezed and measured quadratures is
\begin{equation}
    \theta_j
    =
    \phi_m-\phi_s-\varphi_j,
    \qquad j=0,1 .
    \label{eq:mismatch}
\end{equation}
Including transmission loss and downstream added noise, the variance
projected onto the discrimination quadrature is
\begin{equation}
V_j(r)
=
\frac{\eta_{\rm sq}}{2}
\left[
e^{-2r}\cos^2\theta_j
+
e^{2r}\sin^2\theta_j
\right]
+
\frac{1-\eta_{\rm sq}}{2}
+
V_{\rm add}.
\label{eq:state_variance}
\end{equation}
Here $\eta_{\rm sq}$ is the effective squeezing transmission into the
detected temporal mode, and $V_{\rm add}$ denotes additional noise
referred to the measured quadrature. These effective parameters also
capture propagation and mode-matching imperfections of the detected
squeezed field~\cite{Grimsmo2017,Qiu2023}.

In general, the state-dependent resonator rotations can produce
different projected variances. For the symmetric operating point used
in the baseline analysis, the two effective mismatch angles satisfy
$\theta_0=-\theta_1\equiv\theta_{\rm mis}$, giving
$V_0=V_1\equiv V_{\rm eff}$. We take
$\theta_{\rm mis}=5^\circ$, $\eta_{\rm sq}=1$, and
$V_{\rm add}=0$ for the baseline results, while transmission loss and
added noise are examined in Sec.~\ref{sec:robustness}.

The two $r$-dependent terms in Eq.~(\ref{eq:state_variance}) describe
the competition between squeezed and anti-squeezed fluctuations.
Increasing $r$ initially suppresses the projected noise, whereas at
sufficiently large $r$ the anti-squeezed contribution becomes
dominant for finite $\theta_{\rm mis}$. Consequently,
$V_{\rm eff}$ reaches a finite minimum at $r_{\rm opt}$, as shown in
Fig.~\ref{fig:magnetometer}(c). The full covariance propagation and
the expression for $r_{\rm opt}$ are given in
Appendix~\ref{app:readout}.

\subsection{Quadrature discrimination and assignment error}

The outgoing microwave field is amplified by the cryogenic readout
chain, demodulated with an appropriate phase reference, and integrated
with a temporal weighting chosen to maximize state
distinguishability~\cite{Bultink2018,PhysRevA.86.032106}.
The resulting measurement record is projected onto the discrimination
quadrature $x$ defined above.

The two state-conditioned distributions are Gaussian with means
$m_0$ and $m_1$ and projected variances $V_0$ and $V_1$.
Figures~\ref{fig:magnetometer}(a) and
\ref{fig:magnetometer}(b) compare the corresponding distributions for
vacuum and squeezed readout. The vacuum distributions overlap
substantially, whereas squeezing reduces the projected noise and
thereby suppresses their overlap.

For unequal variances, the optimal assignment boundary follows from
the likelihood-ratio condition. At the symmetric operating point,
$V_0=V_1=V_{\rm eff}$, and the optimal threshold lies midway between
the two pointer-state means. Using the separation defined in
Eq.~(\ref{eq:pointer_separation}), the signal-to-noise ratio and
symmetric assignment error are
\begin{equation}
    {\rm SNR}
    =
    \frac{\Delta m}{\sqrt{V_{\rm eff}}},
    \qquad
    p_{\rm err}
    =
    \frac{1}{2}
    \operatorname{erfc}
    \left(
    \frac{{\rm SNR}}{2\sqrt{2}}
    \right).
    \label{eq:assignment}
\end{equation}

Equation~(\ref{eq:assignment}) connects the projected squeezed noise
directly to state discrimination. At fixed $\Delta m$, reducing
$V_{\rm eff}$ increases the SNR and lowers the assignment error.
Accordingly, $p_{\rm err}$ exhibits the same nonmonotonic dependence
on squeezing as $V_{\rm eff}$ and reaches its minimum at the same
$r_{\rm opt}$, as shown in Fig.~\ref{fig:magnetometer}(c).

We define the binary readout contrast as
$C_{\rm ro}=1-2p_{\rm err}$. Ideal assignment gives
$C_{\rm ro}=1$, while finite assignment error reduces the observed
Ramsey contrast. The readout contrast therefore determines how much
of the Ramsey-encoded magnetic-field information remains accessible
from the assigned binary outcomes, as quantified in Sec.~\ref{sec:fisher}.
The general Gaussian decision rule is given in
Appendix~\ref{app:readout}.

\section{Readout-limited Fisher information and magnetic-field sensitivity}
\label{sec:fisher}

We now quantify how the state-assignment error limits the magnetic-field
information recovered from the Ramsey measurement and how squeezed readout
improves the corresponding magnetometric sensitivity. For the Ramsey sensing
state introduced in Sec.~\ref{sec:magnetometer_model}, the QFI associated
with the magnetic-field variation is
\begin{equation}
F_Q=(G_B T)^2 .
\label{eq:FQ}
\end{equation}
This quantity sets the information limit established by the Ramsey
encoding~\cite{PhysRevA.83.063836,Escher2011,Demkowicz-Dobrzański2012}. Because the final analysis pulse is independent of the unknown field, it preserves the QFI of the Ramsey sensing state. We use this pre-readout QFI as the information benchmark set by the sensing stage; the subsequent readout determines how much of this information is accessible from the observed outcomes.

The information accessible after state assignment is determined by
the observed binary probabilities. Using the readout contrast
$C_{\rm ro}$, the excited-state probability becomes
\begin{equation}
    q_1(B)
    =
    \frac{1}{2}
    \left[
    1-C_{\rm ro}\cos\Phi
    \right],
    \qquad
    \Phi=\phi_B+\phi_a .
    \label{eq:observed_probability}
\end{equation}
The corresponding CFI is
\begin{equation}
    F_C(B)
    =
    \frac{
    [\partial_B q_1(B)]^2
    }{
    q_1(B)[1-q_1(B)]
    } .
    \label{eq:cfi}
\end{equation}
Here $F_C$ quantifies the magnetic-field information retained in the
assigned binary outcomes. Information contained in the continuous
readout record before state assignment is not included in this
quantity.
Substituting Eq.~(\ref{eq:observed_probability}) into
Eq.~(\ref{eq:cfi}) gives
\begin{equation}
    \frac{F_C(B)}{F_Q}
    =
    \frac{
    C_{\rm ro}^2\sin^2\Phi
    }{
    1-C_{\rm ro}^2\cos^2\Phi
    } .
    \label{eq:fi_ratio}
\end{equation}
This ratio quantifies the fraction of the Ramsey-encoded information
that remains accessible after state assignment. For ideal readout,
$C_{\rm ro}=1$ and the population measurement reaches the quantum
limit. Finite assignment error reduces the accessible information and
makes the information recovery dependent on the operating point along
the Ramsey fringe~\cite{PhysRevA.83.063836,Len2022,PRXQuantum.4.040305}.

\begin{figure}[t]
    \centering
    \includegraphics[width=\columnwidth]{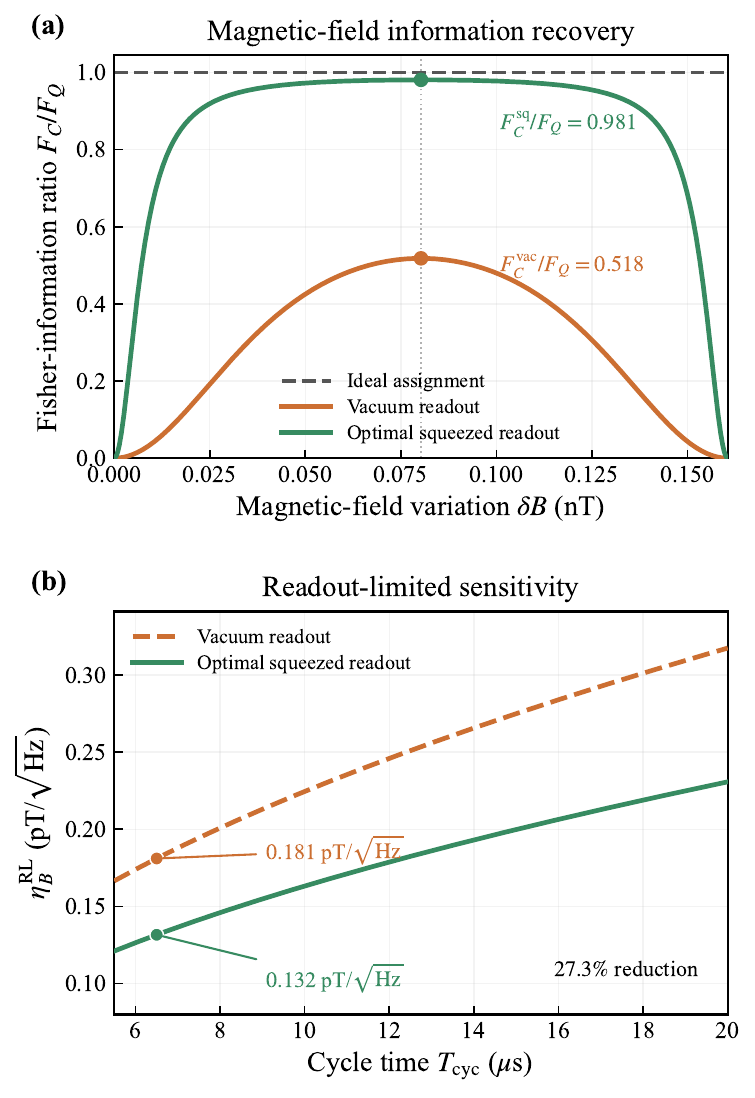}
    \caption{
        \justifying
    Readout-limited Fisher-information recovery and magnetic-field
sensitivity.
(a) $F_C/F_Q$ over one half of the Ramsey fringe for ideal assignment,
vacuum readout, and optimal squeezed readout. The midfringe values are
$0.518$ and $0.981$ for vacuum and squeezed readout, respectively.
(b) Readout-limited sensitivity $\eta_B^{\rm RL}$ versus cycle time
$T_{\rm cyc}$. At $T_{\rm cyc}=6.5~\mu{\rm s}$, optimal squeezing
improves the sensitivity from $0.181$ to
$0.132~{\rm pT}/\sqrt{\rm Hz}$, a $27.3\%$ reduction. }
    \label{fig:fisher_sensitivity}
\end{figure}

Figure~\ref{fig:fisher_sensitivity}(a) shows the accessible Fisher information
over one half of the Ramsey fringe. At the midfringe operating point,
$\Phi=\pi/2$, Eq.~(\ref{eq:fi_ratio}) reduces to
\begin{equation}
    \left.
    \frac{F_C}{F_Q}
    \right|_{\rm mid}
    =
    C_{\rm ro}^2
    =
    (1-2p_{\rm err})^2 .
    \label{eq:fi_ratio_mid}
\end{equation}
For vacuum readout, $p_{\rm err}\simeq0.140$ gives
$F_C/F_Q\simeq0.518$. At the optimal squeezing strength
$r_{\rm opt}\simeq1.22$, the assignment error decreases to
$p_{\rm err}\simeq0.0048$, increasing the accessible fraction to
$F_C/F_Q\simeq0.981$. Thus, optimal squeezed readout recovers nearly
all of the Ramsey-encoded information available from the assigned
binary outcomes.

To translate the recovered Fisher information into magnetometric
performance, we consider $N$ independent Ramsey measurements. For any
locally unbiased estimator of the magnetic field, the Cramér--Rao bound
gives
\begin{equation}
    \delta B
    \ge
    \frac{1}{\sqrt{N F_C}} .
    \label{eq:crb}
\end{equation}
For a total acquisition time
$T_{\rm tot}=N T_{\rm cyc}$, we define the corresponding
Cramér--Rao-limited readout sensitivity benchmark as
\begin{equation}
    \delta B\sqrt{T_{\rm tot}}
    \ge
    \eta_B^{\rm RL}
    \equiv
    \sqrt{\frac{T_{\rm cyc}}{F_C}} .
    \label{eq:readout_sensitivity}
\end{equation}
Thus, $\eta_B^{\rm RL}$ represents the minimum magnetic-field
sensitivity permitted by the Fisher information accessible from the
assigned binary outcomes within the present readout model.
At the Ramsey midfringe, where
$F_C=C_{\rm ro}^2F_Q=C_{\rm ro}^2(G_BT)^2$, this benchmark reduces to
\begin{equation}
    \eta_B^{\rm RL}
    =
    \frac{\sqrt{T_{\rm cyc}}}
    {C_{\rm ro}|G_B|T}.
    \label{eq:readout_sensitivity_mid}
\end{equation}
For fixed Ramsey interrogation and cycle time, the effect of squeezed
readout therefore enters through the readout contrast $C_{\rm ro}$:
improving the state discrimination increases the accessible Fisher
information and lowers the readout-limited sensitivity benchmark.

Figure~\ref{fig:fisher_sensitivity}(b) shows the readout-limited magnetic-field
sensitivity as a function of the cycle time. For the representative
parameters $T=5.0~\mu{\rm s}$, $\tau_m=0.50~\mu{\rm s}$, and
$\tau_{\rm reset}=1.0~\mu{\rm s}$, the cycle time is
$T_{\rm cyc}=6.5~\mu{\rm s}$. At this operating point,
$\eta_B^{\rm RL}$ decreases from approximately
$0.181~{\rm pT}/\sqrt{\rm Hz}$ for vacuum readout to
$0.132~{\rm pT}/\sqrt{\rm Hz}$ with optimal squeezing, corresponding
to a $27.3\%$ reduction. 
Because the squeezed-readout contrast cannot exceed unity, the maximum relative sensitivity improvement for the same Ramsey sequence is \(1-C_{\rm vac}=2p_{\rm err,vac}\simeq28\%\). The obtained reduction therefore approaches the upper bound set by the vacuum-reference assignment error within the present binary-readout model.
The resulting improvement therefore reflects the recovery of
magnetic-field information lost during readout, while the intrinsic
magnetic responsivity and the QFI of the Ramsey sensing state remain
unchanged.

\section{ROBUSTNESS AND EXPERIMENTAL FEASIBILITY}
\label{sec:robustness}

We now examine the robustness of the squeezed-readout enhancement
against experimentally relevant imperfections. We first characterize
its tolerance to quadrature mismatch, then consider squeezing
transmission loss and added measurement noise, and finally discuss
representative device parameters and practical implementation.

\subsection{Robustness to quadrature mismatch}
To quantify the useful operating region, we define the fractional
improvement in the readout-limited magnetic-field sensitivity as
\begin{equation}
    R_B(r,\theta_{\rm mis})
    =
    1-
    \frac{
    \eta_B^{\rm RL}(r,\theta_{\rm mis})
    }{
    \eta_{B,\rm vac}^{\rm RL}
    } ,
    \label{eq:sensitivity_improvement}
\end{equation}
where $\eta_{B,\rm vac}^{\rm RL}$ is the vacuum-readout sensitivity
evaluated with the same Ramsey interrogation and cycle times.
Positive $R_B$ therefore indicates an improvement over vacuum readout,
while $R_B=0$ defines the boundary of the beneficial squeezing regime.

\begin{figure*}[htbp]
    \centering
    \includegraphics[width=0.96\textwidth]{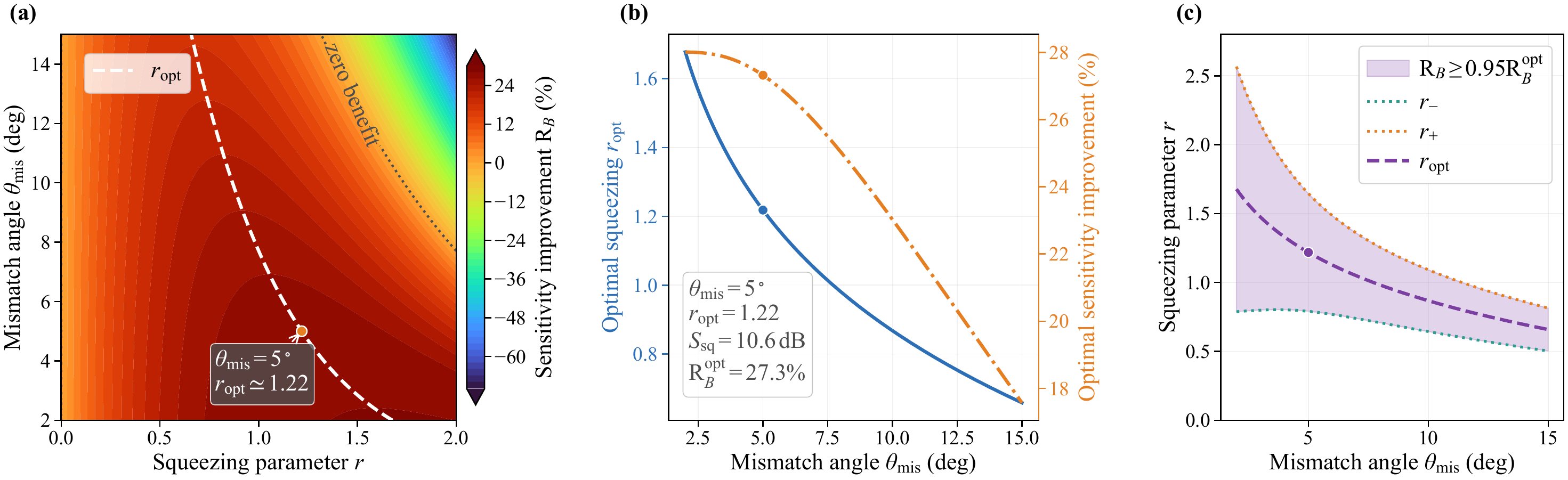}
    \caption{ \justifying
Robustness of the squeezed-readout enhancement to quadrature mismatch.
(a) Readout-limited sensitivity improvement $R_B$ versus squeezing
strength $r$ and mismatch angle $\theta_{\rm mis}$. The white dashed
curve indicates $r_{\rm opt}$, and the dotted contour marks
$R_B=0$. The highlighted point corresponds to
$\theta_{\rm mis}=5^\circ$ and $r_{\rm opt}\simeq1.22$.
(b) Optimal squeezing strength $r_{\rm opt}$ and maximum sensitivity
improvement $R_B^{\rm opt}$ as functions of $\theta_{\rm mis}$.
(c) Near-optimal squeezing window $[r_-,r_+]$ satisfying
$R_B\ge0.95R_B^{\rm opt}$.
}
    \label{fig:robustness}
\end{figure*}

Figure~\ref{fig:robustness}(a) maps $R_B$ over the
$(r,\theta_{\rm mis})$ plane. For small and moderate mismatch angles,
a finite range of squeezing yields a clear improvement over vacuum
readout. At fixed $\theta_{\rm mis}$, $R_B$ first increases with $r$
and then decreases once the projected anti-squeezed noise becomes
dominant~\cite{PhysRevLett.109.153601,PhysRevLett.117.190503,PhysRevApplied.20.054008}.
The resulting
ridge defines the optimal squeezing strength $r_{\rm opt}$, while the
$R_B=0$ contour marks the boundary of the beneficial regime.
Figure \ref{fig:robustness}(b) summarizes how the optimum varies with
the mismatch angle. As $\theta_{\rm mis}$ increases, the optimal
squeezing strength shifts to smaller $r$, while the maximum attainable
sensitivity improvement gradually decreases. For
$\theta_{\rm mis}=5^\circ$, we obtain
$r_{\rm opt}\simeq1.22$,  reproducing the baseline optimum of Sec.~\ref{sec:fisher}.
Figure \ref{fig:robustness}(c) further shows that the optimum is not
confined to a single squeezing strength. We define the near-optimal
window $[r_-,r_+]$ by
\begin{equation}
    R_B(r,\theta_{\rm mis})
    \ge
    0.95\,R_B^{\rm opt}(\theta_{\rm mis}) .
    \label{eq:near_optimal_window}
\end{equation}
The width of this interval gives the tolerance to squeezing-strength
variations. The finite width of this region indicates that the predicted
enhancement does not require precise tuning of the squeezing strength.

\subsection{Robustness to squeezing transmission loss and added noise}
We next examine the impact of imperfect squeezing transmission and
added measurement noise. The mismatch angle is fixed at the
representative value $\theta_{\rm mis}=5^\circ$, while the effective
squeezing transmission $\eta_{\rm sq}$ and added noise $V_{\rm add}$
are varied. For each pair $(\eta_{\rm sq},V_{\rm add})$, the squeezing
strength $r$ is optimized and the corresponding sensitivity
improvement $R_B$ is evaluated using Eq.~(\ref{eq:sensitivity_improvement}).

\begin{figure}[!t]
  \centering
  \includegraphics[width=0.9\linewidth]{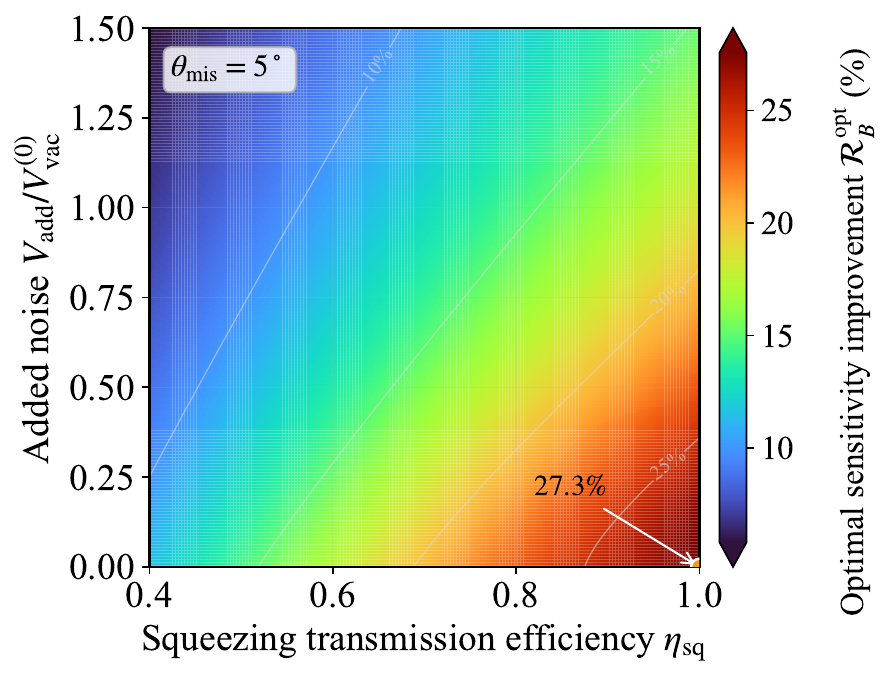}
\caption{\justifying
 Robustness of the squeezed-readout enhancement to squeezing
  transmission loss and added noise. The optimized sensitivity
  improvement $R_B^{\rm opt}$ is shown versus the squeezing
  transmission efficiency $\eta_{\rm sq}$ and normalized added noise
  $V_{\rm add}/V_{\rm vac}^{(0)}$ for
  $\theta_{\rm mis}=5^\circ$. The squeezing strength $r$ is optimized
  at each point. White contours indicate improvements of
  $10\%$, $15\%$, $20\%$, and $25\%$, and the lossless, noiseless
  limit reaches $27.3\%$.
}
\label{fig:loss_noise}
\end{figure}

Figure~\ref{fig:loss_noise} shows the optimized sensitivity improvement
$R_B^{\rm opt}$ as a function of $\eta_{\rm sq}$ and the normalized added noise
$V_{\rm add}/V_{\rm vac}^{(0)}$, with $V_{\rm vac}^{(0)}=1/2$. 
In the lossless and noiseless limit,
$\eta_{\rm sq}=1$ and $V_{\rm add}=0$, the optimized improvement recovers the baseline result of Sec.~\ref{sec:fisher}.
As the squeezing transmission decreases or the added noise increases,
the suppression of the measured-quadrature noise is weakened and the
achievable improvement is progressively reduced.
The degradation remains gradual over a broad range of nonideal
conditions. For example, with no added noise,
$\eta_{\rm sq}=0.8$ still retains an optimal sensitivity improvement
of approximately $23.1\%$. When
$\eta_{\rm sq}=0.8$ and
$V_{\rm add}/V_{\rm vac}^{(0)}=0.5$,
the improvement remains approximately $18.3\%$.

Within the present model, $\eta_{\rm sq}$ and $V_{\rm add}$ reduce the
magnitude of the achievable enhancement without shifting
$r_{\rm opt}$. This is because $\eta_{\rm sq}$ rescales the
$r$-dependent squeezed contribution, while $V_{\rm add}$ enters as an
$r$-independent offset. The optimal squeezing strength therefore
remains determined by the competition between the squeezed and
anti-squeezed quadratures set by $\theta_{\rm mis}$.
More general imperfections, including frequency-dependent loss and
finite squeezing bandwidth, can modify this behavior.
\subsection{Experimental implementation and practical considerations}

The proposed protocol is compatible with a standard circuit-QED
readout architecture, as illustrated in Fig.~\ref{fig:scheme}(a). A coherent readout
tone and a squeezed microwave field are applied to the dispersively
coupled resonator, and the outgoing field is routed through a
circulator and amplified by a low-noise parametric amplifier followed
by a HEMT amplifier~\cite{RevModPhys.93.025005}. After phase-referenced demodulation and temporal integration, the
measurement record is projected onto the calibrated discrimination
quadrature used for state assignment. Representative device and readout parameters are summarized in Table~\ref{tab:parameters}. The vacuum-readout performance is characterized by
${\rm SNR}_{\rm vac}=2.16$, which sets the reference discrimination
performance against which the squeezed-readout enhancement is
evaluated.

The main experimental requirement is to maintain the alignment between
the squeezed quadrature and the discrimination direction throughout
the measurement chain. The squeezing phase should therefore remain
locked to the coherent readout tone~\cite{Lin2014,Lin2013}.
Because the orientation of the detected readout
mode can vary during resonator ring-up and ring-down, the relevant
alignment is determined by the temporally integrated mode. Finite
squeezing bandwidth and frequency-dependent phase shifts can then
produce an effective mismatch $\theta_{\rm mis}$ and reduce the usable
squeezing~\cite{PhysRevLett.106.220502,PhysRevLett.115.093604}.

A further constraint arises at large squeezing strength. Although the
squeezed field has zero coherent amplitude, it carries a mean photon
number
\begin{equation}
    \bar n_{\rm sq}=\sinh^2 r .
    \label{eq:squeezed_photon_number}
\end{equation}
For $r_{\rm opt}=1.218$, this corresponds to approximately
$10.6~{\rm dB}$ of quadrature squeezing and
$\bar n_{\rm sq}\simeq2.38$. The increased fluctuation power can raise
the intracavity photon population and enhance measurement-induced
transitions or other non-QND effects outside the ideal dispersive
regime~\cite{PhysRevLett.109.153601,PhysRevLett.117.190503,PhysRevApplied.22.064038}. The practical operating point must therefore balance readout-noise
suppression against quadrature mismatch, transmission loss, finite
bandwidth, and measurement backaction.
\begin{table}[t]
\caption{Representative sensor and readout parameters used in the
numerical calculations.}
\label{tab:parameters}
\centering
\begin{ruledtabular}
\begin{tabular}{lc}
Quantity & Value \\
\hline

\multicolumn{2}{c}{\textit{Transmon sensor}} \\[2pt]
$E_C/h$                         & $0.25~{\rm GHz}$ \\
$E_{J,\max}/h$                  & $20~{\rm GHz}$ \\
$d$                             & $0.08$ \\
$A_{\rm eff}$                   & $100~\mu{\rm m}^2$ \\
$\Phi_{\rm bias}/\Phi_0$        & $0.35$ \\
$f_{01}$                        & $4.0200~{\rm GHz}$ \\
$\partial_B f_{01}$             & $-0.6228~{\rm MHz/nT}$ \\[3pt]

\multicolumn{2}{c}{\textit{Dispersive readout}} \\[2pt]
$\kappa/2\pi$                   & $1.50~{\rm MHz}$ \\
$\epsilon_d/2\pi$               & $0.81~{\rm MHz}$ \\
$\chi_0/2\pi$                   & $-0.687~{\rm MHz}$ \\
$\chi_1/2\pi$                   & $+0.687~{\rm MHz}$ \\
$\Delta/2\pi$                   & $0~{\rm MHz}$ \\
$\tau_m$                        & $0.50~\mu{\rm s}$ \\[3pt]

\multicolumn{2}{c}{\textit{Readout model and timing}} \\[2pt]
${\rm SNR}_{\rm vac}$           & $2.16$ \\
$\theta_{\rm mis}$              & $5^\circ$ \\
$\eta_{\rm sq}$                 & $1$ \\
$V_{\rm add}/V_{\rm vac}^{(0)}$ & $0$ \\
$T$                             & $5.0~\mu{\rm s}$ \\
$\tau_{\rm reset}$              & $1.0~\mu{\rm s}$ \\

\end{tabular}
\end{ruledtabular}
\end{table}

\section{CONCLUSION}
\label{sec:conclusion}
We have developed an effective detected-mode framework for superconducting-qubit magnetometry with squeezed dispersive readout, quantitatively connecting projected quadrature noise, state-assignment error, the classical Fisher information accessible from binary readout outcomes, and the resulting readout-limited magnetic-field sensitivity. Squeezing improves the sensing performance by reducing information loss during state assignment while leaving the magnetic responsivity and the QFI established during Ramsey interrogation unchanged. A finite mismatch between the squeezed quadrature and the discrimination direction leads to an optimal squeezing strength through the competition between squeezed and anti-squeezed fluctuations. For the representative operating point considered here, the accessible Fisher-information fraction increases from $0.518$ to $0.981$, while the readout-limited magnetic-field sensitivity is improved by $27.3\%$, approaching the upper bound set by the vacuum-reference assignment error within the present binary-readout model. The enhancement persists over a finite squeezing window and remains robust against quadrature mismatch, transmission loss, and added measurement noise. These results establish a quantitative link between squeezed-microwave readout and measurement-stage information recovery and provide a practical framework for assessing squeezing-assisted readout in superconducting quantum sensing.

\begin{acknowledgments}
This
work was supported by the National Natural Science Foundation of China under Grants No. 12274376, No. U21A20434,
No. 12074346, and No. 12575032; the High-level Talent
Research Start-up Project Funding of Henan Academy of
Sciences (Project No. 242027151); and the Fundamental Research Fund of Henan Academy of Sciences (Project No.
20261827003).
\end{acknowledgments}

\section*{DATA AVAILABILITY}
	The data supporting the findings of this article are publicly available~\cite{yun_2026_22669959}. 

\appendix

\section{Flux-dependent transmon model}
\label{app:transmon}
For completeness, we summarize the flux-dependent transmon model
used to determine the magnetic transduction coefficient in
Eq.~(\ref{eq:magnetic_transduction}).

We denote by $\Phi_{\rm ext}$ the total external magnetic flux
threading the SQUID loop. For a SQUID with junction asymmetry $d$,
the effective Josephson energy is
\begin{equation}
E_J(\Phi_{\rm ext})
=
E_{J,\max}
\sqrt{
\cos^2\left(\frac{\pi\Phi_{\rm ext}}{\Phi_0}\right)
+
d^2\sin^2\left(\frac{\pi\Phi_{\rm ext}}{\Phi_0}\right)
}.
\end{equation}
The corresponding transmon Hamiltonian is

\begin{equation}
H_q
=
4E_C(\hat n-n_g)^2
-
E_J(\Phi_{\rm ext})\cos\hat{\varphi}.
\end{equation}

We take $\Phi_{\rm bias}$ to denote the static external flux at the
selected magnetic-field operating point $B_0$. A small field variation
$\delta B=B-B_0$ changes the external flux according to

\begin{equation}
\Phi_{\rm ext}(B)
=
\Phi_{\rm bias}
+
A_{\rm eff}\delta B,
\end{equation}
where $A_{\rm eff}$ is the effective loop area.

Numerical diagonalization of $H_q$ gives the field-dependent qubit
transition frequency
$f_{01}(B)=\omega_q(B)/(2\pi)$.
Around the selected operating point $B_0$, the angular transition
frequency can be linearized as

\begin{equation}
\omega_q(B_0+\delta B)
\simeq
\omega_q(B_0)
+
\left.
\frac{\partial\omega_q}{\partial B}
\right|_{B_0}
\delta B .
\end{equation}
Using the magnetic transduction coefficient defined in
Eq.~(\ref{eq:magnetic_transduction}), this can equivalently be written as

\begin{equation}
\omega_q(B_0+\delta B)
\simeq
\omega_q(B_0)
+
G_B\delta B .
\end{equation}

For the parameters listed in Table~\ref{tab:parameters}, we obtain

\begin{equation}
\left.
\frac{\partial f_{01}}{\partial B}
\right|_{B_0}
=
\frac{G_B}{2\pi}
\simeq
-0.62~\mathrm{MHz/nT}.
\end{equation}

\section{Dispersive readout and state-dependent squeezed-noise model}
\label{app:readout}

In this Appendix, we provide the detailed readout model underlying
Eqs.~(\ref{eq:cavity_response})--(\ref{eq:assignment}) of the main
text. We first derive the state-conditioned resonator response and
the corresponding scattering of incident fluctuations. We then
propagate the squeezed covariance to the detected temporal mode and
derive the Gaussian assignment error used in the main text.

\subsection{Dispersive resonator response}

We consider a transmon coupled to a single microwave readout
resonator. Before making the dispersive approximation, the
qubit--resonator system is described by the Jaynes--Cummings
Hamiltonian
\begin{equation}
\frac{H}{\hbar}
=
\omega_r a^\dagger a
+
\frac{\omega_q}{2}\sigma_z
+
g\left(a^\dagger\sigma_-+a\sigma_+\right)
+
\frac{H_d(t)}{\hbar},
\label{eq:B_JC}
\end{equation}
where $a$ is the resonator annihilation operator, $\omega_r$ is the
bare resonator frequency, $\omega_q$ is the qubit transition
frequency, and $g$ is the qubit--resonator coupling strength. The
coherent readout drive is
\begin{equation}
\frac{H_d(t)}{\hbar}
=
\epsilon_d(t)a^\dagger e^{-i\omega_d t}
+
\epsilon_d^*(t)a e^{i\omega_d t},
\label{eq:B2}
\end{equation}
with drive frequency $\omega_d$ and complex envelope $\epsilon_d(t)$.

In the dispersive regime,
\begin{equation}
|\omega_q-\omega_r|\gg g,
\end{equation}
the exchange interaction can be eliminated perturbatively, giving
\begin{equation}
\frac{H_{\rm disp}}{\hbar}
=
(\omega_r+\chi\sigma_z)a^\dagger a
+
\frac{\widetilde{\omega}_q}{2}\sigma_z
+
\frac{H_d(t)}{\hbar}.
\label{eq:B_disp}
\end{equation}
Conditioned on the qubit state $|j\rangle$, with
$j\in\{0,1\}$, the resonator Hamiltonian in the frame rotating at
$\omega_d$ becomes
\begin{equation}
\frac{H_j}{\hbar}
=
(\Delta+\chi_j)a^\dagger a
+
\epsilon_d(t)a^\dagger
+
\epsilon_d^*(t)a,
\label{eq:B_conditional}
\end{equation}
where $\Delta=\omega_r-\omega_d$. For the symmetric dispersive
convention, $\chi_0=-\chi$ and $\chi_1=+\chi$.

Including resonator damping at rate $\kappa$, the corresponding quantum
Langevin equation~\cite{PhysRevA.31.3761} is
\begin{equation}
\dot a
=
-
\left[
\frac{\kappa}{2}
+
i(\Delta+\chi_j)
\right]a
-i\epsilon_d(t)
+
\sqrt{\kappa}\,a_{\rm in}(t),
\label{eq:B7}
\end{equation}
where $a_{\rm in}(t)$ denotes the zero-mean incident fluctuations.
Taking the expectation value and defining
$\alpha_j(t)=\langle a(t)\rangle_j$ gives
\begin{equation}
\dot{\alpha}_j(t)
=
-\left[
\frac{\kappa}{2}
+i(\Delta+\chi_j)
\right]\alpha_j(t)
-i\epsilon_d(t),
\label{eq:B_alpha}
\end{equation}
which reproduces Eq.~(\ref{eq:cavity_response}) of the main text.

For a constant readout drive, the steady-state amplitude is
\begin{equation}
\alpha_j^{\rm ss}
=
\frac{-i\epsilon_d}
{\kappa/2+i(\Delta+\chi_j)}.
\label{eq:B_alpha_ss}
\end{equation}
The state dependence of $\alpha_j$ generates the coherent
pointer-state separation used for qubit discrimination.

\subsection{State-dependent scattering and detected temporal mode}
The outgoing field is related to the intracavity and incident fields
through the input--output relation
\begin{equation}
a_{\rm out}(t)
=
a_{\rm in}(t)-\sqrt{\kappa}\,a(t).
\label{eq:B_input_output}
\end{equation}
Because $\langle a_{\rm in}(t)\rangle=0$, its state-conditioned mean is
\begin{equation}
\langle a_{\rm out}(t)\rangle_j
=
-\sqrt{\kappa}\,\alpha_j(t).
\label{eq:B_output_mean}
\end{equation}
Thus, within the linear dispersive model, the zero-mean squeezed
field modifies the output fluctuations without changing the coherent
mean generated by the readout drive.

To determine how the fluctuations propagate through the resonator,
we write
\begin{equation}
a(t)=\alpha_j(t)+\delta a_j(t).
\end{equation}
The fluctuation operator satisfies
\begin{equation}
\delta\dot a_j
=
-\left[
\frac{\kappa}{2}
+i(\Delta+\chi_j)
\right]\delta a_j
+
\sqrt{\kappa}\,a_{\rm in}(t).
\label{eq:B_fluctuation}
\end{equation}
In the frequency domain,
\begin{equation}
\delta a_j[\omega]
=
\sqrt{\kappa}\,
\chi_{c,j}[\omega]\,
a_{\rm in}[\omega],
\end{equation}
with the state-conditioned cavity susceptibility
\begin{equation}
\chi_{c,j}[\omega]
=
\frac{1}
{\kappa/2+i(\Delta+\chi_j-\omega)}.
\label{eq:B_susceptibility}
\end{equation}
The corresponding output fluctuations are therefore
\begin{equation}
\delta a_{{\rm out},j}[\omega]
=
S_j(\omega)a_{\rm in}[\omega],
\label{eq:B_scattering_general}
\end{equation}
where
\begin{equation}
S_j(\omega)
=
1-\kappa\chi_{c,j}[\omega].
\label{eq:B_Somega}
\end{equation}

For squeezing narrowband compared with the relevant resonator
response, $S_j(\omega)$ may be evaluated near the readout frequency,
giving
\begin{equation}
S_j
=
1-
\frac{\kappa}
{\kappa/2+i(\Delta+\chi_j)}
\equiv e^{i\varphi_j}.
\label{eq:B_state_scattering}
\end{equation}
For an ideal lossless single-port resonator, $|S_j|=1$, so the
state-dependent resonator response acts as a phase rotation
$\varphi_j$ of the incident fluctuations. Equation
(\ref{eq:B_state_scattering}) gives the state-dependent scattering
relation used in Eq.~(\ref{eq:state_scattering}) of the main text.

For a finite squeezing bandwidth, the frequency dependence of
$S_j(\omega)$ remains inside the temporal-mode integration and can
produce additional phase dispersion and mode mismatch. These effects,
together with propagation loss and imperfect mode matching, are
represented below by effective detected-mode parameters.

The continuous output field is integrated with a normalized temporal
mode $f(t)$,
\begin{equation}
Z
=
\int_0^{\tau_m}dt\,f(t)a_{\rm out}(t),
\qquad
\int_0^{\tau_m}dt\,|f(t)|^2=1.
\label{eq:B_temporal_mode}
\end{equation}
Here $\tau_m$ is the readout integration time. We introduce the
canonical quadratures of this temporal mode,
\begin{equation}
X=\frac{Z+Z^\dagger}{\sqrt{2}},
\qquad
P=\frac{Z-Z^\dagger}{i\sqrt{2}},
\label{eq:B_quadratures}
\end{equation}
and collect them in
$\mathbf R=(X,P)^T$.

The binary readout considered in the main text uses a single
calibrated quadrature,
\begin{equation}
x
=
\mathbf u_m^T\mathbf R,
\qquad
\mathbf u_m=
\begin{pmatrix}
\cos\phi_m\\
\sin\phi_m
\end{pmatrix},
\label{eq:B_measured_quadrature}
\end{equation}
where $\phi_m$ specifies the discrimination direction. The two
projected means are
\begin{equation}
m_j=\langle x\rangle_j,
\label{eq:B_projected_mean}
\end{equation}
and their separation is the $\Delta m$ defined in
Eq.~(\ref{eq:pointer_separation}) of the main text.

Equation~(\ref{eq:B_quadratures}) is used only to represent the
phase-space covariance of the detected temporal mode. The assignment
model itself depends only on the projected measurement variable $x$.

\subsection{Propagation of the squeezed covariance}
For a squeezed vacuum with squeezing strength $r$, the covariance
matrix in its principal-axis frame is
\begin{equation}
\Sigma_0(r)
=
\frac{1}{2}
\begin{pmatrix}
e^{-2r} & 0\\
0 & e^{2r}
\end{pmatrix},
\label{eq:B_sigma0}
\end{equation}
with the vacuum limit
\begin{equation}
\Sigma_{\rm vac}
=
\frac{1}{2}\mathbb I.
\end{equation}
For an input squeezing angle $\phi_s$, the covariance before the
state-dependent resonator response is
\begin{equation}
\Sigma_{\rm in}(r,\phi_s)
=
R(\phi_s)\Sigma_0(r)R^T(\phi_s),
\label{eq:B_sigma_in}
\end{equation}
where
\begin{equation}
R(\phi)
=
\begin{pmatrix}
\cos\phi & -\sin\phi\\
\sin\phi & \cos\phi
\end{pmatrix}.
\end{equation}

In the narrowband lossless limit, the resonator phase
$\varphi_j$ rotates the covariance according to
\begin{equation}
\Sigma_j^{\rm cav}
=
R(\varphi_j)
\Sigma_{\rm in}(r,\phi_s)
R^T(\varphi_j).
\label{eq:B_state_covariance}
\end{equation}
The two qubit states therefore generally produce different output
covariances.

We model transmission loss between the squeezed source and the
detected temporal mode by an effective efficiency $\eta_{\rm sq}$.
Unsqueezed vacuum admixed by loss and downstream added noise then give
\begin{equation}
\Sigma_{j}^{\rm det}
=
\eta_{\rm sq}\Sigma_j^{\rm cav}
+
(1-\eta_{\rm sq})\frac{\mathbb I}{2}
+
V_{\rm add}\mathbb I.
\label{eq:B_detected_covariance}
\end{equation}
The variance of the measured quadrature is
\begin{equation}
V_j
=
\mathbf u_m^T
\Sigma_j^{\rm det}
\mathbf u_m.
\label{eq:B_projected_variance}
\end{equation}

Defining the effective post-resonator mismatch angle
\begin{equation}
\theta_j
=
\phi_m-\phi_s-\varphi_j,
\label{eq:B_theta}
\end{equation}
Eq.~(\ref{eq:B_projected_variance}) becomes
\begin{equation}
V_j(r)
=
\frac{\eta_{\rm sq}}{2}
\left[
e^{-2r}\cos^2\theta_j
+
e^{2r}\sin^2\theta_j
\right]
+
\frac{1-\eta_{\rm sq}}{2}
+
V_{\rm add},
\label{eq:B_Vj}
\end{equation}
which reproduces Eq.~(\ref{eq:state_variance}) of the main text.

Equation~(\ref{eq:B_Vj}) makes explicit that the two
state-conditioned readout distributions need not have the same
variance. State-dependent resonator rotation, frequency-dependent
filtering, or asymmetric mode matching can all generate
$V_0\neq V_1$.

For the symmetric baseline used in the main text,
$\Delta=0$ and $\chi_0=-\chi_1$. In the narrowband single-port
limit the corresponding reflection coefficients satisfy the
conjugate symmetry $S_1=S_0^*$. With the squeezing phase chosen
symmetrically about the discrimination direction, the effective
mismatch angles satisfy
\begin{equation}
\theta_0=-\theta_1\equiv\theta_{\rm mis}.
\label{eq:B_symmetric_theta}
\end{equation}
Because Eq.~(\ref{eq:B_Vj}) depends only on
$\cos^2\theta_j$ and $\sin^2\theta_j$, the projected variances then
coincide,
\begin{equation}
V_0=V_1\equiv V_{\rm eff}.
\label{eq:B_symmetric_variance}
\end{equation}

For the baseline value $\theta_{\rm mis}=5^\circ$, the competition
between the squeezed and anti-squeezed terms produces a finite minimum
of $V_{\rm eff}$. Minimizing $V_{\rm eff}$ with respect to $r$ gives
\begin{equation}
e^{4r_{\rm opt}}
=
\cot^2\theta_{\rm mis},
\end{equation}
and therefore
\begin{equation}
r_{\rm opt}
=
\frac{1}{2}
\ln\!\left(\cot|\theta_{\rm mis}|\right),
\qquad
0<|\theta_{\rm mis}|<\frac{\pi}{4}.
\label{eq:B_ropt}
\end{equation}
For $\theta_{\rm mis}=5^\circ$, this gives
$r_{\rm opt}\simeq1.22$. Constant transmission loss and isotropic
added noise change the attainable minimum variance but do not shift
$r_{\rm opt}$ within this effective model. Frequency-dependent loss,
finite-bandwidth filtering, and state-dependent mode mismatch can
modify this simple result.

\subsection{Gaussian discrimination and assignment error}

After projection onto the calibrated discrimination quadrature, the
state-conditioned measurement distributions are
\begin{equation}
P(x|j)
=
\frac{1}{\sqrt{2\pi V_j}}
\exp\left[
-\frac{(x-m_j)^2}{2V_j}
\right].
\label{eq:B_gaussian}
\end{equation}

For equal prior probabilities, the minimum-error decision rule follows
from the likelihood-ratio condition. The boundary between the two
conditional distributions satisfies
\begin{equation}
P(x|0)=P(x|1),
\label{eq:B_likelihood_boundary}
\end{equation}
or equivalently
\begin{equation}
\frac{(x-m_0)^2}{V_0}
-
\frac{(x-m_1)^2}{V_1}
=
\ln\!\left(\frac{V_1}{V_0}\right).
\label{eq:B_general_threshold}
\end{equation}
For unequal variances, this equation determines the relevant
likelihood-ratio crossing or crossings, and the corresponding
assignment probabilities follow by integrating the two Gaussian
distributions over their decision regions.

At the symmetric operating point,
$V_0=V_1=V_{\rm eff}$, the likelihood-ratio boundary reduces to the
midpoint
\begin{equation}
x_{\rm th}
=
\frac{m_0+m_1}{2}.
\label{eq:B_threshold}
\end{equation}
The two conditional assignment errors are then equal,
\begin{equation}
p_{0\rightarrow1}
=
p_{1\rightarrow0}
\equiv p_{\rm err},
\end{equation}
and evaluating the Gaussian tail gives
\begin{equation}
p_{\rm err}
=
\frac{1}{2}
\operatorname{erfc}
\left(
\frac{\Delta m}
{2\sqrt{2V_{\rm eff}}}
\right).
\label{eq:B_perr}
\end{equation}
With
\begin{equation}
{\rm SNR}
=
\frac{\Delta m}{\sqrt{V_{\rm eff}}},
\label{eq:B_SNR}
\end{equation}
this becomes
\begin{equation}
p_{\rm err}
=
\frac{1}{2}
\operatorname{erfc}
\left(
\frac{{\rm SNR}}{2\sqrt{2}}
\right),
\label{eq:B_perr_SNR}
\end{equation}
which is Eq.~(\ref{eq:assignment}) of the main text.

The complete readout chain used in the baseline calculation can
therefore be summarized as
\begin{equation}
S_j
\longrightarrow
\theta_j
\longrightarrow
V_j
\longrightarrow
V_{\rm eff}
\longrightarrow
{\rm SNR}
\longrightarrow
p_{\rm err}.
\label{eq:B_readout_chain}
\end{equation}
The symmetric reduction $V_0=V_1=V_{\rm eff}$ is specific to the
baseline operating point. The general state-dependent formulation
above allows asymmetric projected variances to be incorporated.

\bibliographystyle{apsrev4-2}
\bibliography{references}

\end{document}